\documentclass[a4paper,12pt]{article}
\usepackage{jcappub}
\usepackage{amsthm,graphicx}
\usepackage{epsfig}
\usepackage{latexsym, amssymb} 
\usepackage{amsmath}
\def\beq{\begin{equation}}
\def\eeq{\end{equation}}
\def\br{\begin{eqnarray}}
\def\er{\end{eqnarray}}
\def\benu{\begin{enumerate}}
\def\efnu{\end{enumerate}}
\def\nn{\nonumber}

\def\cl{{\cal C}_{\ell}}

\begin{document}
\title{Confronting the concordance model of cosmology with Planck data}
 \author[a]{Dhiraj Kumar Hazra} 
 \author[a,b]{Arman Shafieloo}

\affiliation[a]{Asia Pacific Center for Theoretical Physics, Pohang, Gyeongbuk 790-784, Korea}
\affiliation[b]{Department of Physics, POSTECH, Pohang, Gyeongbuk 790-784, Korea}

\emailAdd{dhiraj@apctp.org, arman@apctp.org}

\abstract 
{We confront the concordance (standard) model of cosmology, the spatially flat $\Lambda$CDM Universe with power-law 
form of the primordial spectrum with Planck CMB angular power spectrum data searching for possible smooth 
deviations beyond the flexibility of the standard model. The departure from the concordance cosmology is 
modeled in the context of Crossing statistic and statistical significance of this deviation is used as a 
measure to test the consistency of the standard model to the Planck data. 
Derived Crossing functions suggest the presence of some broad features in angular spectrum beyond the expectations of the concordance model. 
Our results indicate that the concordance model of cosmology is consistent to the Planck data only at 2 to 3$\sigma$ confidence level 
if we allow smooth deviations from the angular power spectrum given by the concordance model. 
This might be due to random fluctuations or may hint towards smooth features in the primordial spectrum or departure from another aspect of the standard model.
Best fit Crossing functions indicate that there are lack of power in the data at both low-$\ell$ and high-$\ell$ with respect to the concordance model. This hints that we may 
need some modifications in the foreground modeling to resolve the significant inconsistency at high-$\ell$. However, presence of 
some systematics at high-$\ell$ might be another reason for the deviation we found in our analysis.

}

\maketitle

\section{Introduction}
The assumption of Friedmann-Lema\^{\i}tre-Robertson-Walker spatially flat Universe with a nearly scale invariant primordial spectrum along with cosmological constant as dark 
energy, non-relativistic dark matter (cold dark matter) and baryons as the constituents of the universe is till date the 
most popular model of cosmology, commonly referred to as the concordance $\Lambda$CDM cosmology. 
Recent Planck data~\cite{Planck:2013kta,Ade:2013zuv} has indicated that assuming the standard $\Lambda$CDM model, the mean values of some of the main cosmological parameters need to be shifted from the values we knew in pre-Planck era, especially from WMAP CMB data~\cite{Hinshaw:2012fq}. Two questions arise here, first, if the two main CMB data, WMAP and Planck are consistent with each other considering the fact that they are observing the same Universe, and second, if the standard model of cosmology is the correct model to address the observational data. 
In~\cite{Hazra:2013oqa} we have confronted WMAP and Planck data within the framework of $\Lambda$CDM and we have argued that WMAP and Planck data are consistent with each other in the multipole range covered by WMAP if we allow an overall shift in the amplitudes of the power spectra (for a different analysis see~\cite{Spergel:2013rxa}). In~\cite{Hazra:2013oqa} we have shown that if we fix the amplitudes of the angular power spectra to their reported values by Planck and WMAP surveys, the two observations are inconsistent (worse than $3\sigma$ confidence limit). In this paper we try to address the second question. We use Crossing statistic~\cite{Shafieloo:2010xm,Shafieloo:2012jb,Shafieloo:2012yh,Shafieloo:2012pm} to check the consistency of the concordance model of cosmology with the recent Planck data. In this approach one can test consistency of a model to a given data by comparing it with its own variations. Variations that are substantially preferred by the data can in fact point towards the necessity of a venture beyond the expectations of the
standard model. This novel approach has been shown to be efficient to deal with the data with complicated correlations~\cite{Shafieloo:2012jb}. In this work we use the likelihood code provided by Planck team and we show that there are in fact some variations that can have substantially better likelihood to the Planck data than the best fit standard model and we discuss about statistical significance of these findings in details.   

The paper is organized as follows. In the next section we shall briefly discuss the Crossing statistic and the way to incorporate the method to check the 
consistency between the standard model and the observational data. Next, we present our results and conclude at the end. 

\section{Formalism}~\label{sec:formalism}
 

 The standard model of cosmology stands today as a theory which can be described by 6 parameters. Baryon density ($\Omega_{\rm b}$), cold dark matter 
 density ($\Omega_{\rm CDM}$), the Hubble parameter ($H_0$) and the reionization optical depth ($\tau$) are the background parameters which describe
 the constituents and the expansion history of the Universe. The primordial perturbations are modeled by the spectral amplitude of scalar perturbation $A_{\rm S}$ and its tilt $n_{\rm s}$. 
Using recent cosmological observations including Planck CMB data the parameters of the concordance model are constrained with a great precision,
 however, there still remains a fundamental question. How do we know that the standard model is the actual model of the Universe? This question is 
 usually approached by comparing the predictions of the standard model within its flexibilities with other proposed models or models based on the 
 standard model with some further extensions (hence having more degrees of freedom). In such approaches the most we can do is to clarify 
 which model is favored by the data while none of the proposed  models might be the actual true model. 

While considering observational uncertainties one may never be able to find out the actual model of the Universe~\cite{Shafieloo:2011zv}, we can nevertheless try to 
falsify the consistency of different models, such as the standard model of cosmology, with observations. Using Crossing statistic ~\cite{Shafieloo:2012jb,Shafieloo:2012yh,Shafieloo:2012pm} one can in fact falsify a cosmological model without comparing it with other models or assume any particular form of parametrization for cosmological quantities (see~\cite{Aghamousa:2011kj,Aghamousa:2012bk} as another model independent approach to estimate the $\cl$ directly from the data). 
In this approach a model is compared with its own variations. A Crossing function that is usually based on an orthogonal basis, models the 
deviation from a proposed hypothesis. Depending on the significance of the likelihood improvements to the data coming from the Crossing function, 
one can argue about consistency of the proposed model to the data. In this approach to describe a data (it can be correlated or un-correlated)
we use a modified function which has two parts. One part is given by the theoretical model which provides us with the mean function and another 
part is the Crossing function that models the deviation from the mean function. Mean function (assumed theoretical model) can have different 
forms for each set of its parameters and the Crossing function also can have different shapes for each set of its hyperparameters. 
Statistical analysis are hence based on confidence limits of the parameters of the mean function and hyperparameters of the Crossing function. 
In this formalism and in this work we fit the Planck CMB data with $ {\cal C}_{\ell}^{\rm TT}\mid_{\rm modified}^{N}$ as shown in Eq.\ref{eq:main}. The first part of this modified 
function, ${\cal C}_{\ell}^{\rm TT}\mid_{\rm \Omega_{\rm b}, \Omega_{\rm CDM}, H_0, \tau, A_{\rm s}, n_{\rm s}}$ provides us with the pool of mean
function realizations, which are theoretical ${\cal C}_{\ell}$ from the standard model of cosmology and the second part $T_{N}(C_0,C_1,C_2,...,C_N,\ell)$ is 
the Crossing function which models the deviation from the expectations of the standard model.
 
\beq
 {\cal C}_{\ell}^{\rm TT}\mid_{\rm modified}^{N} ={\cal C}_{\ell}^{\rm TT}\mid_{\rm \Omega_{\rm b}, \Omega_{\rm CDM}, H_0, \tau, A_{\rm s}, n_{\rm s}}~\times ~ T_{N}(C_0,C_1,C_2,...,C_N,\ell).
\label{eq:main} 
\eeq
If the actual mean function (for Crossing function equal to one) can result to a good likelihood fitting the data, in comparison to the modified variations, we can conclude that the proposed model can describe the data pretty well and no modification is required. However, if we realize that including a smooth form of a Crossing function can substantially improve the fit to the data then we can argue that the given mean function might not be a good choice to explain the observational data. In this work following previous papers~\cite{Shafieloo:2012jb,Shafieloo:2012yh,Shafieloo:2012pm,Hazra:2013oqa} we use Chebyshev polynomials as Crossing functions. Orthogonality and smoothness of Chebyshev polynomials and their convergence in a limited range makes them a suitable choice as a Crossing function. Below, we provide the Crossing functions up to fifth order (till the order discussed in this paper): 
\begin{eqnarray}
T_{\rm 0}(C_0,x)&=&C_0 \nn\\
T_{\rm I}(C_0,C_1,x)&=&T_{\rm 0}(C_0,x)+C_1~x\nn\\
T_{\rm II}(C_0,C_1,C_2,x)&=&T_{\rm I}(C_0,C_1,x)+C_2(2x^2-1)\nn\\
T_{\rm III}(C_0,C_1,C_2,C_3,x)&=&T_{\rm II}(C_0,C_1,C_2,x)+C_3(4x^3-3x)\nn\\
T_{\rm IV}(C_0,C_1,C_2,C_3,C_4,x)&=&T_{\rm III}(C_0,C_1,C_2,C_3,x)+C_4(8x^4-8x^2+1)\nn\\
T_{\rm V}(C_0,C_1,C_2,C_3,C_4,C_5,x)&=&T_{\rm IV}(C_0,C_1,C_2,C_3,C_4,x)+C_5(16x^5-20x^3+5x).~~~~~\label{eq:Crossing-function}  
\end{eqnarray}
In our case $x=\ell/\ell_{\rm max}$, where $\ell_{\rm max}=2500$. Hyperparameters of these Crossing functions are the discriminators between true and false models. 
Marginalizing over all mean function parameters we can compute the confidence contours. Consistency of a model to a given data is estimated by measuring the distance of this confidence ball from the position of the mean function described by $C_0=1, C_1=C_2=...=C_N=0$ point in the hyperparameter space.

To begin with we consider the best fit $\Lambda$CDM angular power spectrum as a single mean 
function realization and probe the hyperparameter space of the Crossing functions. Next we allow all the cosmological
parameters to vary, generating pool of possible mean function realizations, along with varying the Crossing hyperparameters. This allow us to understand 
where (in the marginalized likelihood space) the concordance model of cosmology is standing when it is compared with its own variations. Due to orthonormality,
 a higher order Crossing function cannot be mimicked by its lower orders. Step by step we consider higher orders of Crossing functions 
allowing more flexible deviation from the concordance model. Marginalizing over the mean function parameters (cosmological parameters) the location of the confidence contours of the Crossing hyperparameters indicate if more complicated deviations from the assumed mean function are required or not. 

 We have used publicly available software CAMB~\cite{cambsite,Lewis:1999bs} to generate the angular power spectrum from the concordance model and 
 CosmoMC~\cite{cosmomcsite,Lewis:2002ah} with Planck likelihood code to perform the Markov Chain Monte Carlo (MCMC) analysis. 
 Along with the mean function parameters and Crossing hyperparameters we do also marginalize over the 14 nuisance parameters corresponding to 
different foreground and calibration effects in different frequency channels. We should mention that as we are not using any polarization data, we 
have used a Gaussian prior on the reionization optical depth $\tau$.

\section{Results and discussions}\label{sec:results}

\begin{figure*}[!htb]
\begin{center} 
\resizebox{360pt}{240pt}{\includegraphics{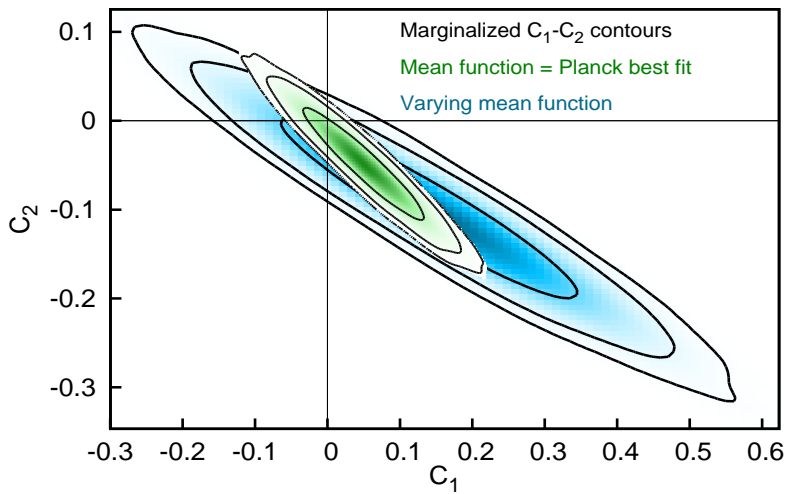}} 
\end{center}
\caption{\footnotesize\label{fig:cntrsplancktoplanck} The confidence contours of Crossing hyperparameters $C_1$ and $C_2$ are plotted.
Here we used second order Crossing function. The green contours correspond to the case that we used the best fit concordance model
(to the Planck data) as the only realization of the mean function. The blue contours represent the case that we allow the parameter of the mean function to vary.
The intersection of two straight black lines represent the position of the concordance model in the hyperparameter space. In both the cases the concordance model is 
just $1\sim2\sigma$ away from the center of the contours.}
\end{figure*}

In Fig.~\ref{fig:cntrsplancktoplanck} we plot the obtained 3$\sigma$ marginalized confidence
contours of $C_1-C_2$ hyperparameters using Crossing function of second order. The green contours represent the case that the best 
fit concordance model is used as a single realization of the mean function. In this case we just vary the Crossing hyperparameters and the Planck foreground nuisance parameters. 
The concordance model is indicated by the intersection of the two black straight lines at $C_1=C_2=0$. Blue contours represent the case where along with Crossing hyperparameters 
we also vary all mean function parameters (cosmological parameters associated to the concordance model). Having larger degrees of freedom the size of the blue contour is substantially 
larger than the green one. One can clearly see that considering Crossing function of the second order, the concordance model is standing at the edge of 1$\sigma$ CL. This shows that 
by considering the Crossing function of the second order the concordance model is still pretty much consistent to the data. We can also notice that the axis of the 
Crossing hyperparameter contours changes slightly when we use different realizations of the mean function. This shows that there are different combinations of the mean function and the
Crossing function (for different sets of parameters and hyperparameters) that can result to a good fit to the data.

\begin{figure*}[!htb]
\begin{center} 
\resizebox{230pt}{160pt}{\includegraphics{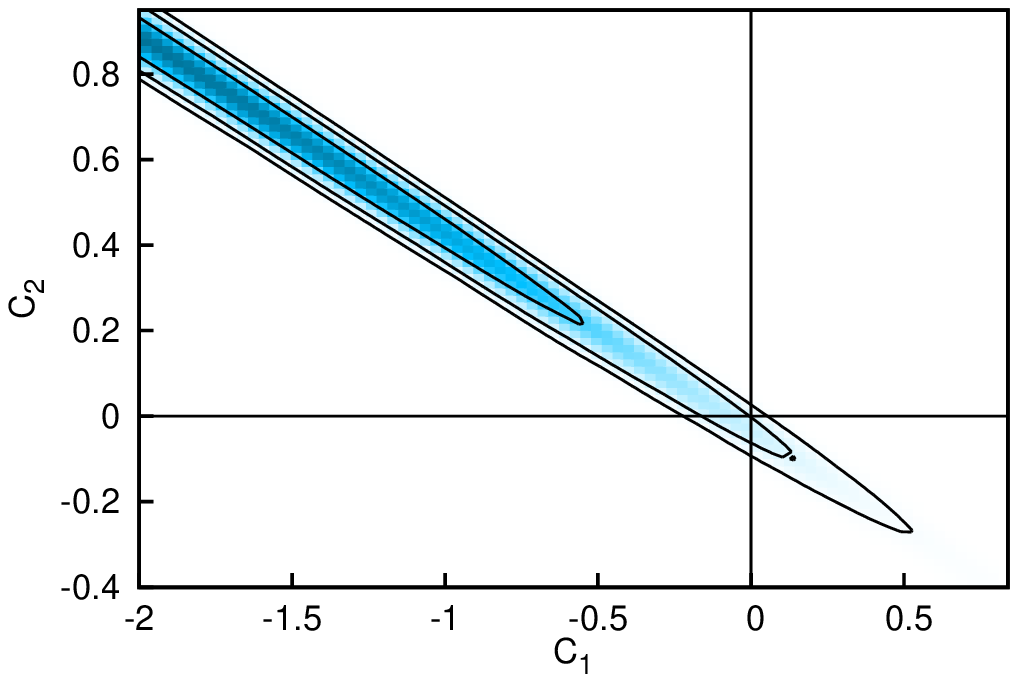}} 
\hskip -30pt \resizebox{230pt}{160pt}{\includegraphics{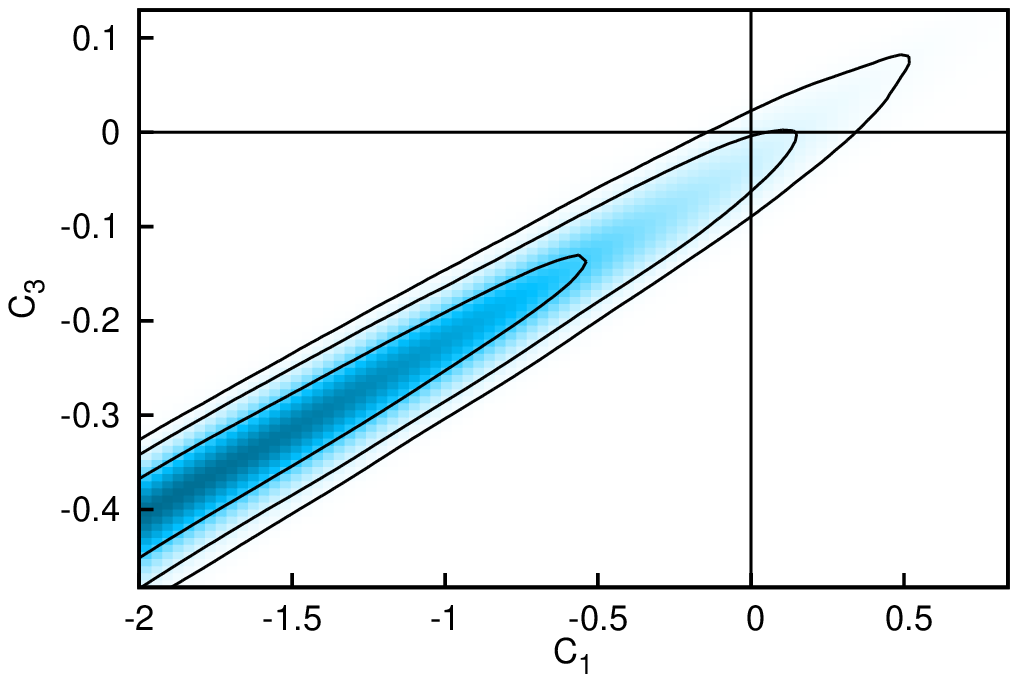}} 
\resizebox{230pt}{160pt}{\includegraphics{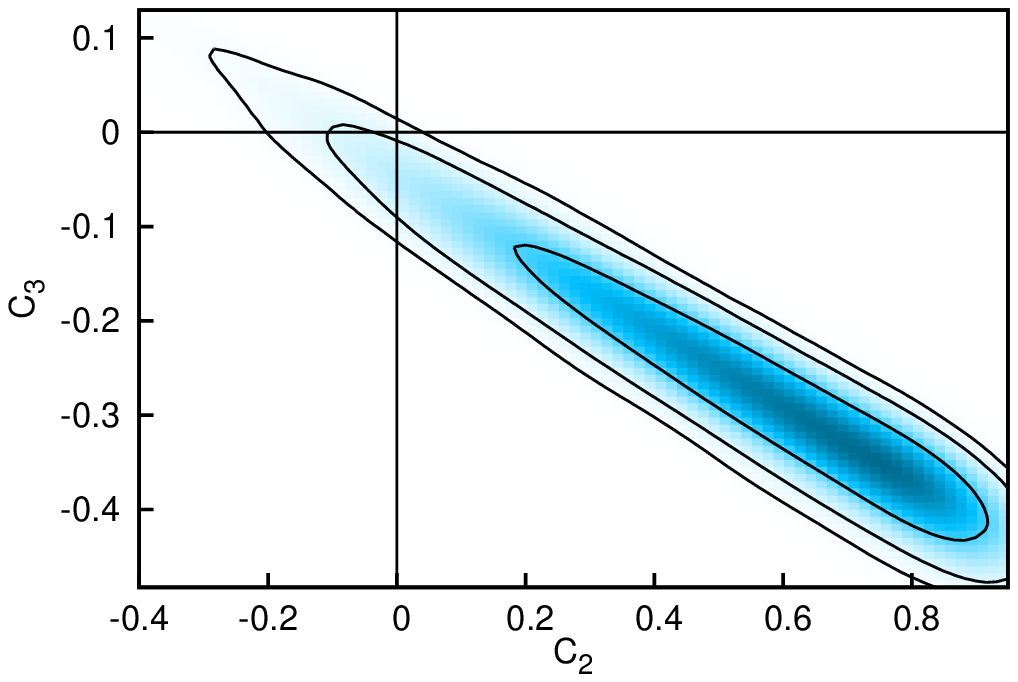}} 
\end{center}
\caption{\footnotesize\label{fig:cntrsplancktoplanckc3} Confidence contours of $C_1,C_2$ and $C_3$ hyperparameters are plotted 
using the third order Crossing function. Here we allowed the parameters of the mean function (concordance model) to vary and results are 
obtained by marginalizing over $C_0$ and all mean function parameters. The position of the concordance model that is shown as the
intersection of the two straight black lines (corresponding to {\it zero} modification by the Crossing function) is now outside 
2$\sigma$ CL. This indicates considerable inconsistency between the concordance model and the Planck data.}
\end{figure*}

As the second order Crossing function may not supply sufficient and appropriate additional flexibilities to the concordance model,
we increase the Crossing level to third order following Eq.~\ref{eq:Crossing-function}. Results are shown in Fig.~\ref{fig:cntrsplancktoplanckc3}. 
Here we only provide the result of the analysis where we allow the parameters of the mean function (concordance model) to vary. 
We should note that allowing the parameters of the mean function to vary is a more conservative approach in testing consistency of a model to the data compared to the
case where we use only a single best fit realization of the mean function. As we can see in Fig.~\ref{fig:cntrsplancktoplanckc3}, by implementing the third 
order Crossing function, the concordance model is pushed outside 2$\sigma$ confidence level. These results suggest that some features beyond the flexibilities 
of the concordance model might be preferred by the Planck data. 

\begin{figure*}[!htb]
\begin{center} 
\resizebox{200pt}{140pt}{\includegraphics{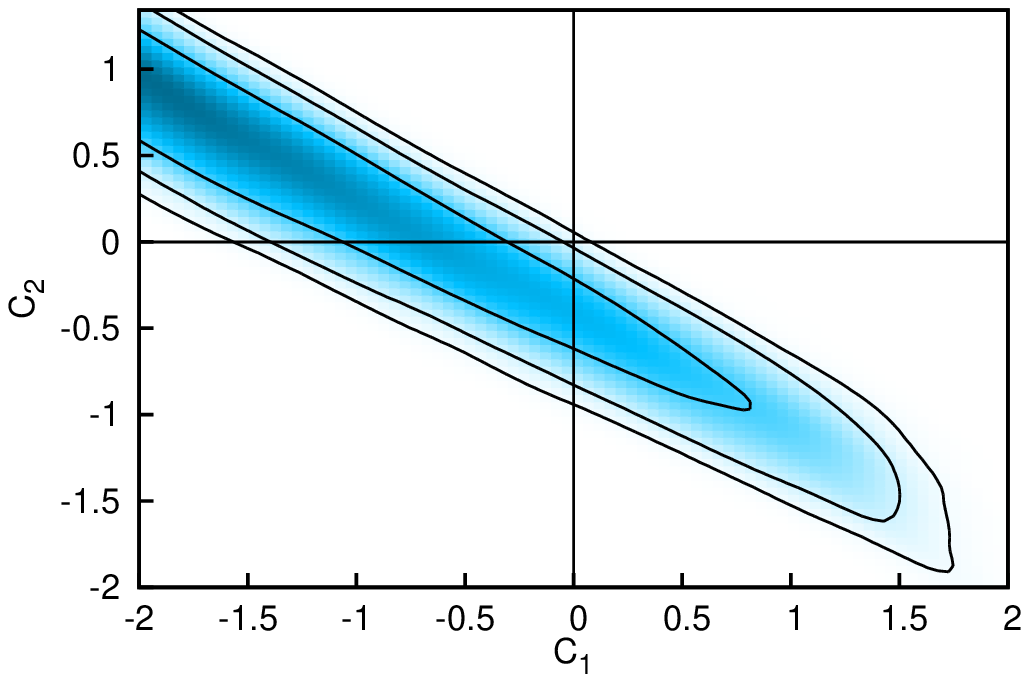}} 
\resizebox{200pt}{140pt}{\includegraphics{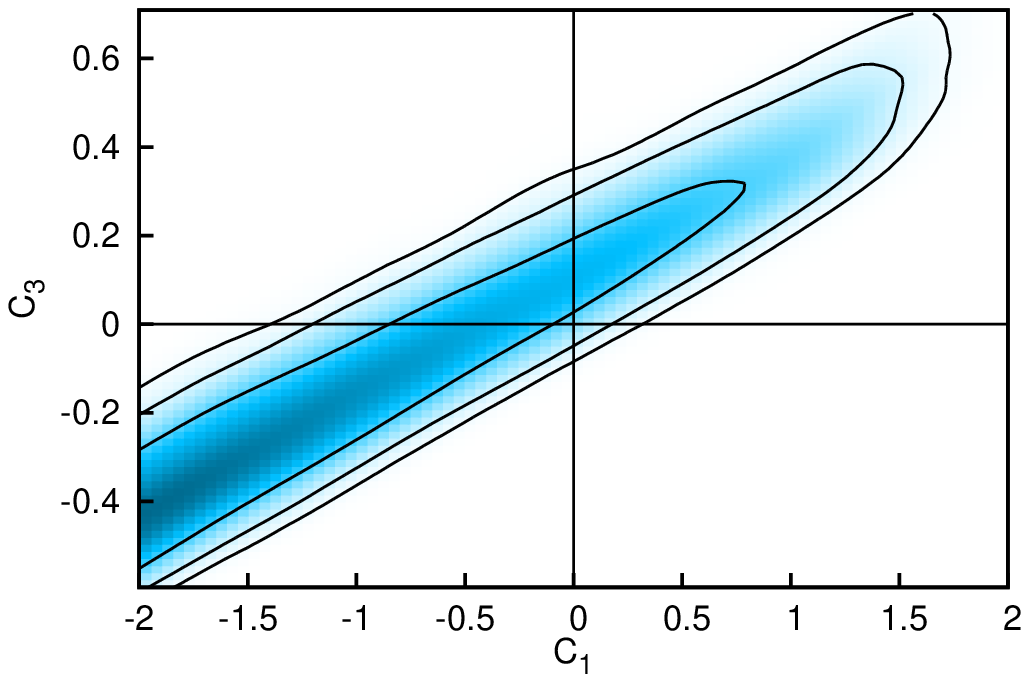}} 
\resizebox{200pt}{140pt}{\includegraphics{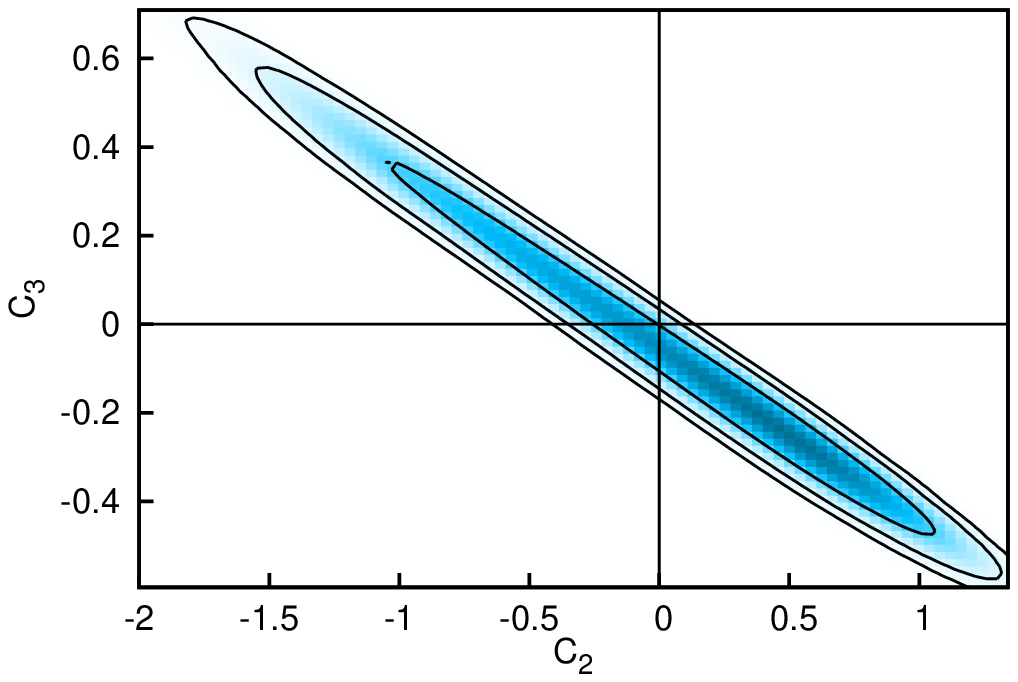}} 
\resizebox{200pt}{140pt}{\includegraphics{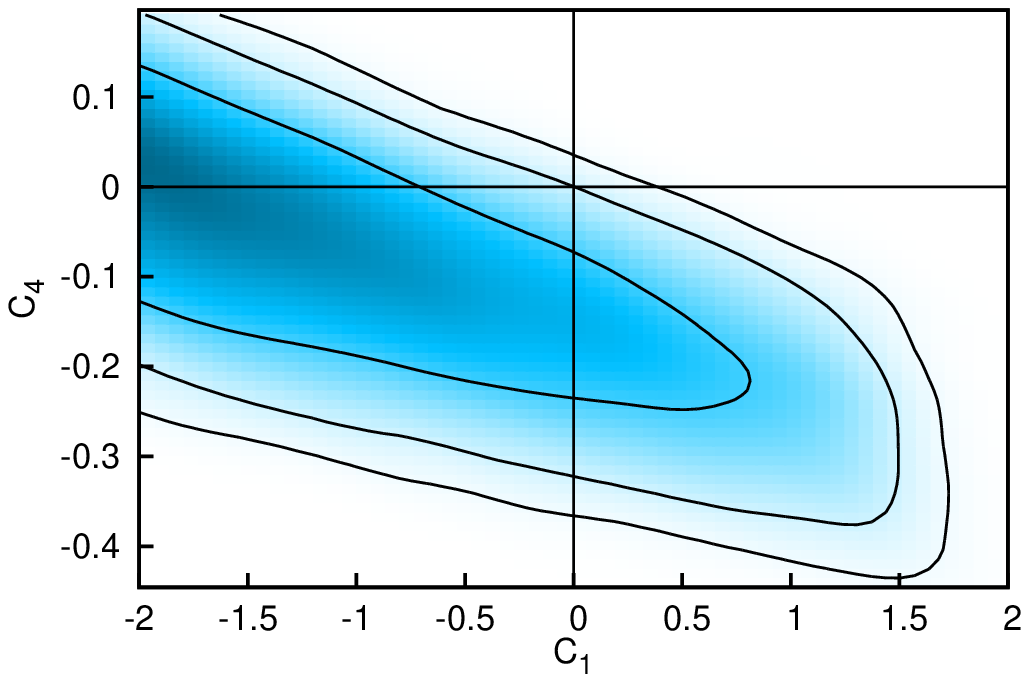}} 
\resizebox{200pt}{140pt}{\includegraphics{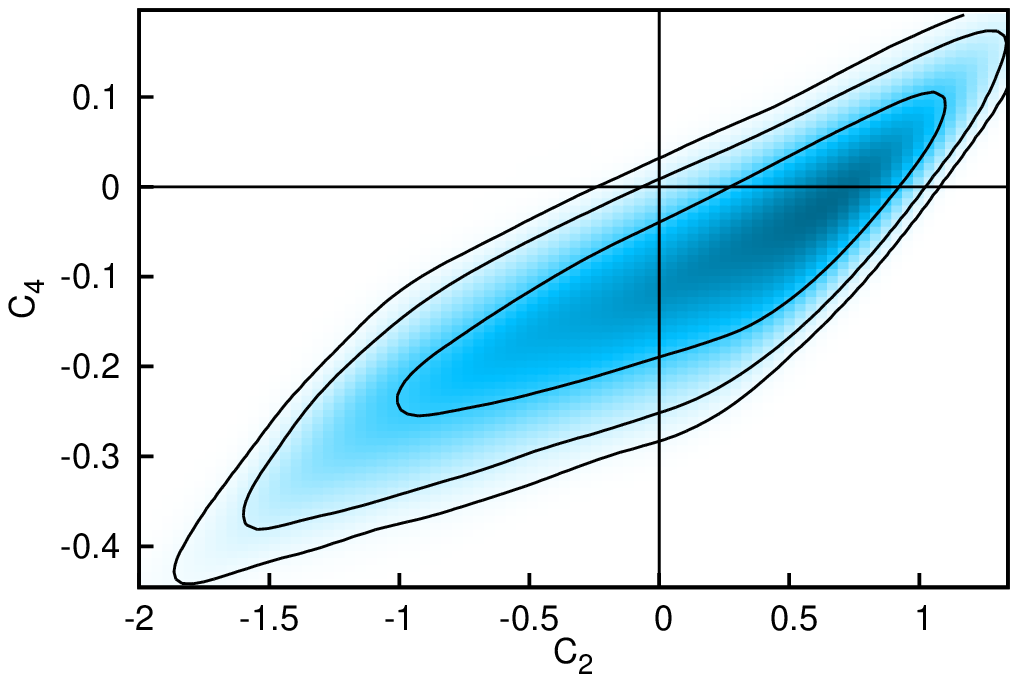}} 
\resizebox{200pt}{140pt}{\includegraphics{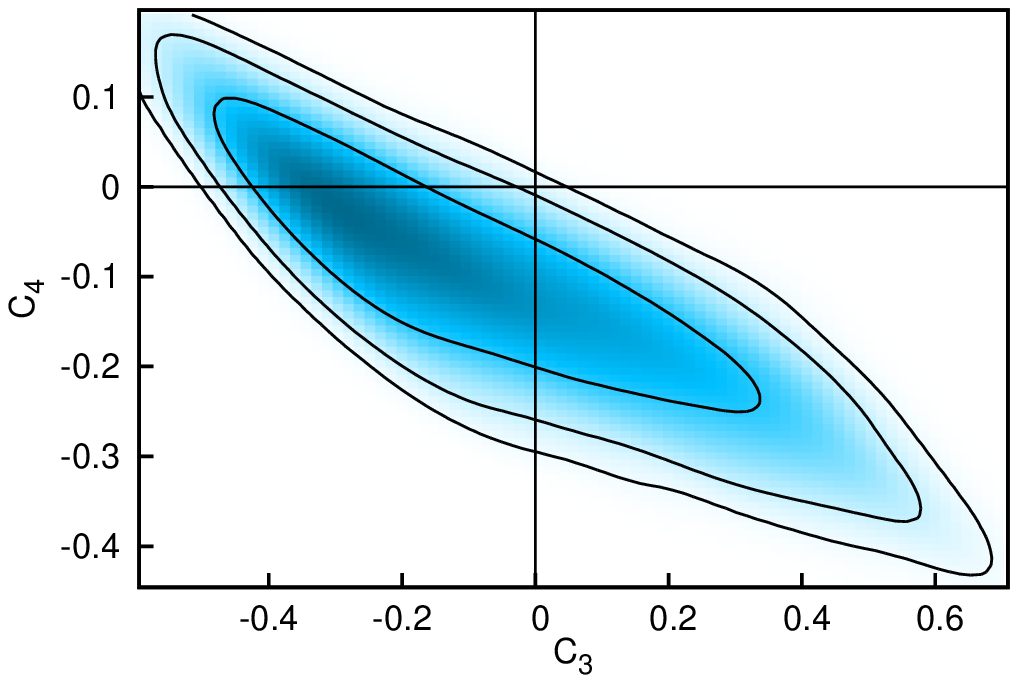}} 
\end{center}
\caption{\footnotesize\label{fig:cntrsplancktoplanckc4} Confidence contours of $C_1,C_2, C_3$ and $C_4$ hyperparameters 
are plotted using the fourth order Crossing function. Here we allowed the parameters of the mean function to vary and results are obtained by marginalizing 
over $C_0$ and all cosmological parameters of the concordance model (parameters of the mean function). The position of the concordance model that is shown 
as the intersection of the straight black lines (corresponding to {\it zero} modification by the Crossing function) is 
outside 2$\sigma$ CL. Results here are similar to Fig.~\ref{fig:cntrsplancktoplanckc3}.}
\end{figure*}

Having found that implementing the third order Crossing function can significantly improve the likelihood to the data, we 
test if adding more flexibility to the Crossing function by going to its fourth order can improve further the fit to the data. 
Results for fourth order Crossing function are provided in Fig.~\ref{fig:cntrsplancktoplanckc4}. Here too we allow the the parameters of the mean 
function to vary. Looking at these figures we do not see much difference from Fig.~\ref{fig:cntrsplancktoplanckc3} and we still find 
the concordance model between 2$\sigma$ and 3$\sigma$ confidence limits of the Crossing hyperparameters. However, due to increase in 
degeneracies by including the fourth order term in Eq.~\ref{eq:Crossing-function} the size of the confidence contours are slightly increased. 

Implementing higher order Crossing functions, the concordance model may find itself back inside the confidence contours of the hyperparameters (due to further increase in degeneracies), 
unless, similar to the case of the third order Crossing function, we happen to find another particular form of flexibility in higher order Crossing functions that fits the data 
substantially better. We have performed the analysis implementing fifth order Crossing function and we did not find considerably different results from 
the fourth and third order cases. 
Beyond the fifth order we did not perform our analysis since a complete MCMC with higher degrees of freedom becomes computationally expensive and extremely time consuming. 



\begin{figure*}[!htb]
\begin{center} 
\resizebox{220pt}{150pt}{\includegraphics{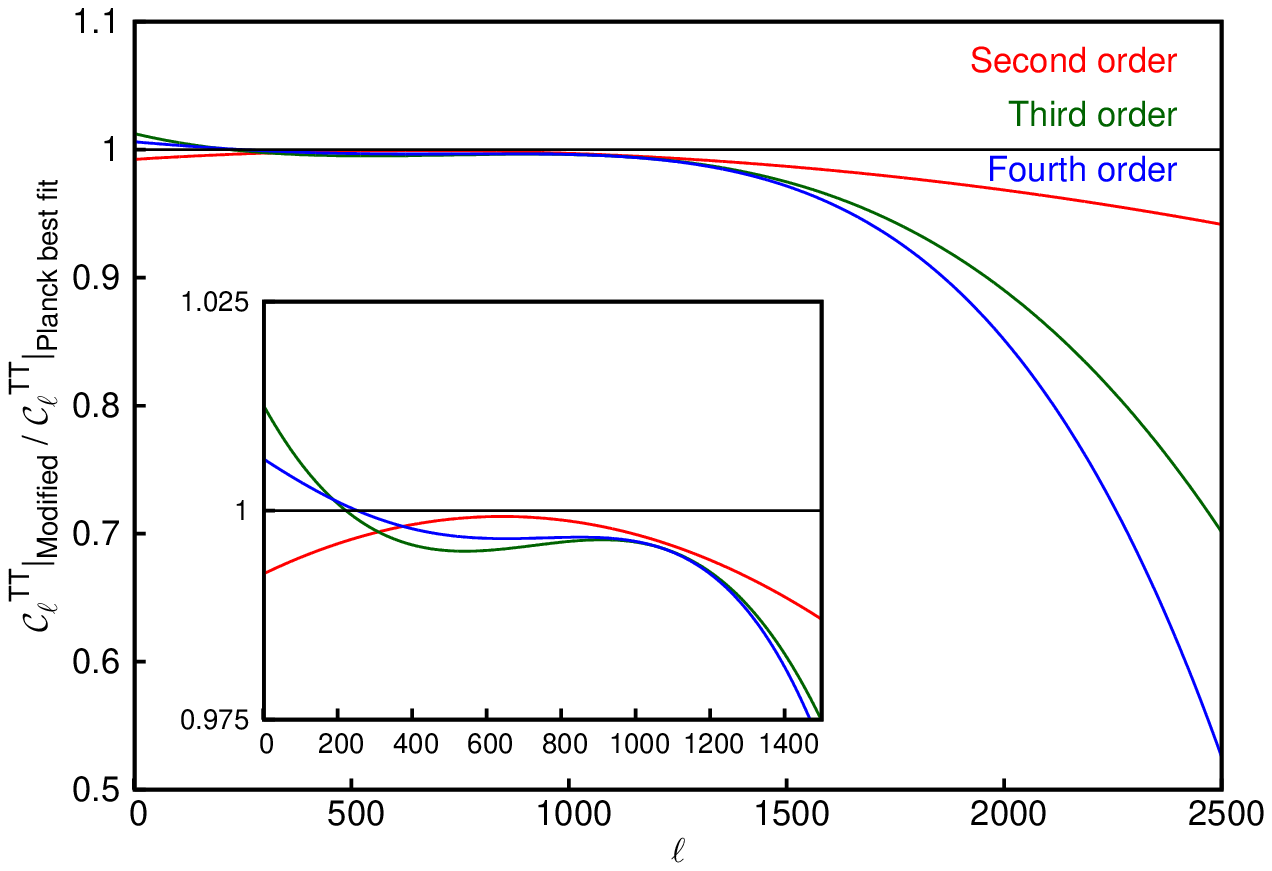}} 
\hskip -8pt \resizebox{220pt}{150pt}{\includegraphics{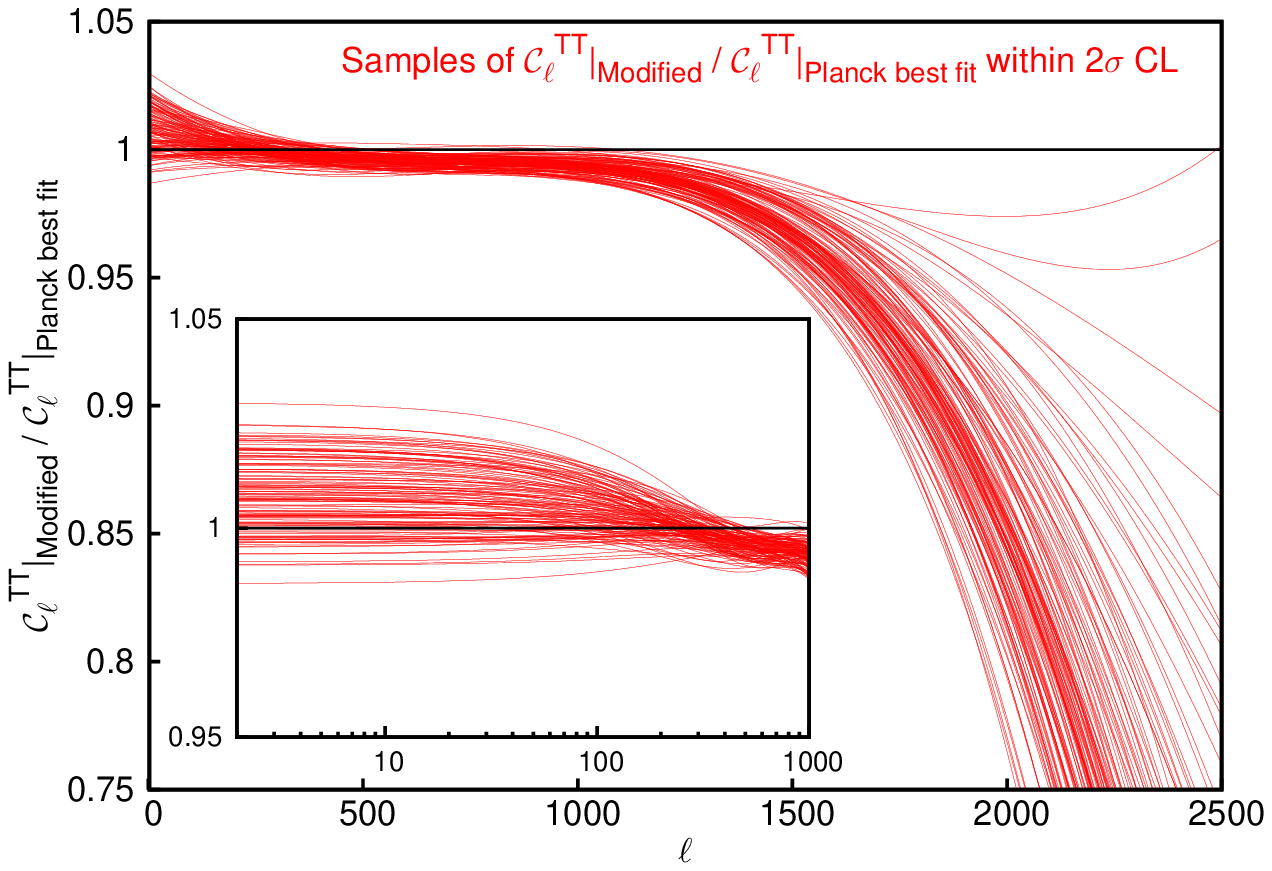}}
\resizebox{220pt}{150pt}{\includegraphics{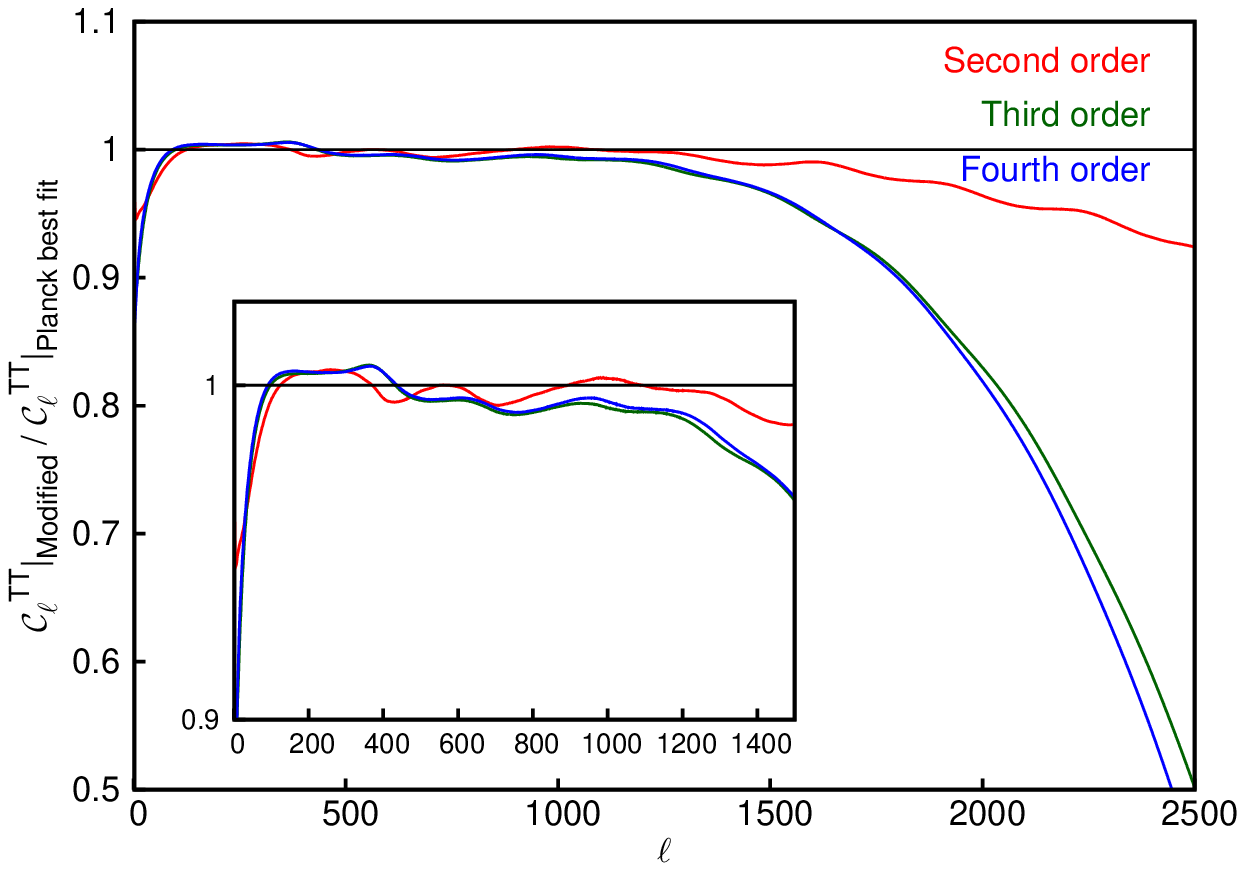}} 
\hskip -8pt \resizebox{220pt}{150pt}{\includegraphics{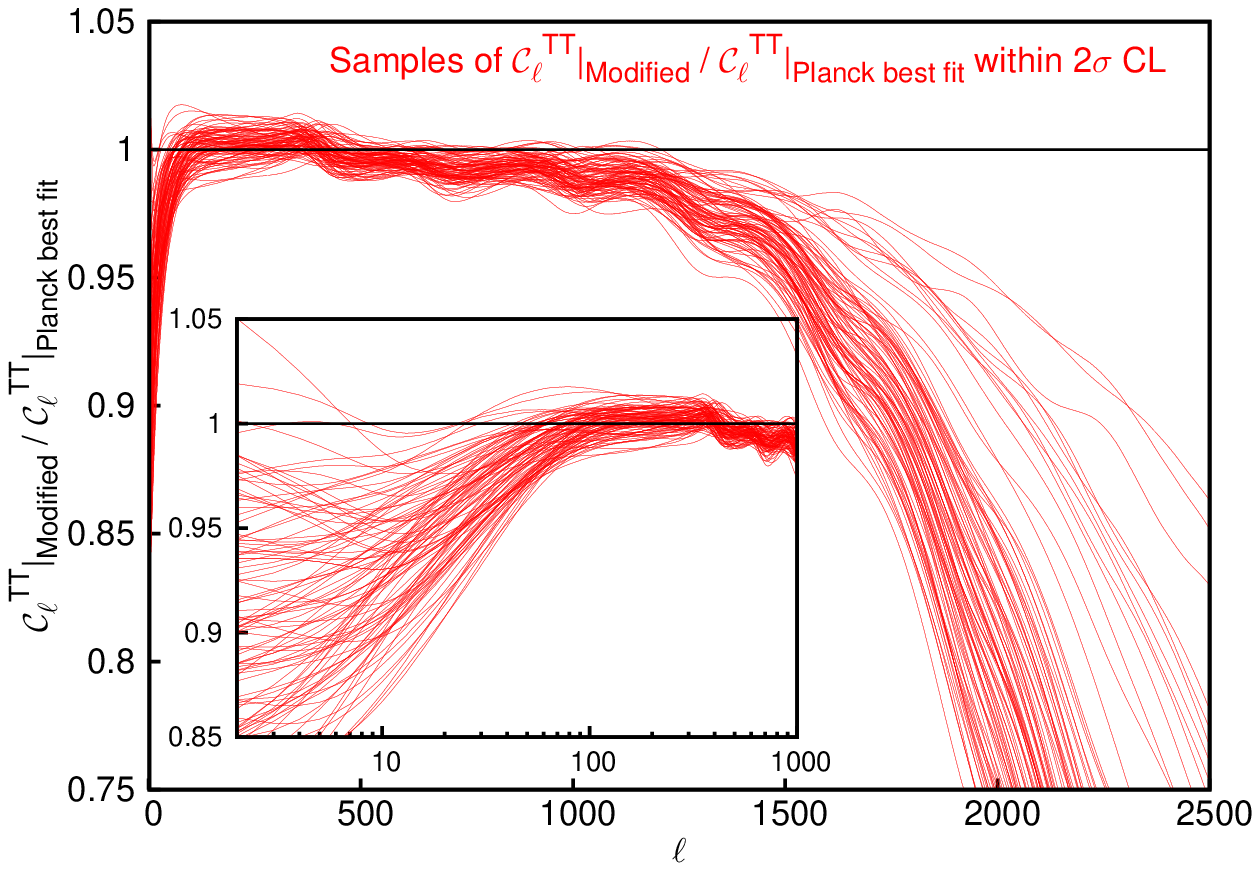}}
\end{center}
\caption{\footnotesize\label{fig:cf} Top panels: The best fit modifications (left) and a non-exhaustive sample of modifications (right) with respect to the best fit concordance model for different orders of Crossing function are plotted. In top panels the best fit concordance model is used as the only realization of the mean function. Bottom panels: similar to the top panels but here we allowed the parameters of the mean function to vary. Crossing functions suggest that there are significant lack of power in the data at both low-$\ell$ and high-$\ell$ with respect to the concordance model.}
\end{figure*}

In left panels of Fig.~\ref{fig:cf} we have plotted the best fit modifications to the concordance model suggested by
Crossing functions of different orders. In the top-left panel the mean function is considered to be fixed at the best fit concordance
model and in the bottom-left panel we allowed the parameters of the mean function to vary. In the right panels we can see a non-exhaustive samples of Crossing modifications (within $2\sigma$ CL from the best fit combination) with respect to the best fit concordance model to the Planck data. Results shown in lower panels indicate that implementing the Crossing functions of different orders, we may find other points in the parameter space of the  mean function (rather than the best fit concordance model) more suitable to fit the Planck data. Constant horizontal line at one represent the best fit concordance model. All these samples shown in these plots have substantially better likelihood to the Planck data with respect to the best fit concordance model.    

Looking at these results we can see that a damping at low multipoles and also a damping at high multipoles (with respect to the theoretical angular power 
spectrum given by the concordance model) are suggested by the data. We should note that lack of power in the data at low multipoles (in comparison 
to the concordance model) is well known, e.g. see~\cite{Planck:2013kta,Hazra:2013nca}. In our analysis we model this departure and show quantitatively how much this deviation from the concordance model is statistically significant. Our analysis indicates that while at low multipoles we may have a lack 
of the power in the data (with respect to concordance model), at smaller scales too (near $\ell=2000-2500$) the Crossing functions suggest a significant damping
in the power spectrum. At these scales different foreground effects dominate over the effects from primordial fluctuations and the background Universe 
and this may suggest a requirement for some modifications in the foreground modeling. We should note here that $N_{eff}$ for neutrinos is fixed to $3.046$, the standard model value. Following Planck analysis the neutrino mass follows the standard mass hierarchy and mass of the single massive neutrino eigenstate is assumed to be $m_{\nu}=0.06$ eV throughout the analysis.

\begin{figure*}[!htb]
\begin{center} 

\hskip -8pt \resizebox{420pt}{300pt}{\includegraphics{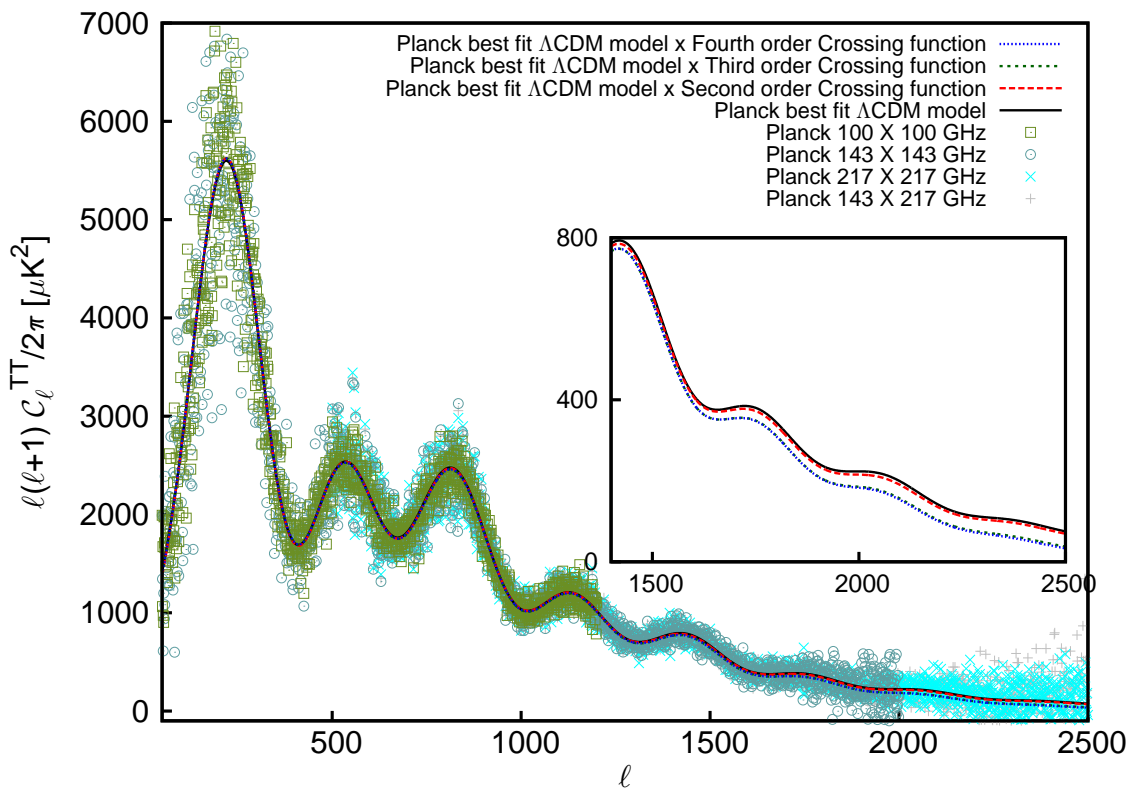}} 
\end{center}
\caption{\footnotesize\label{fig:clttplanck}  Planck data in different frequencies and the best fit $\cl$'s obtained implementing different orders of the Crossing function are plotted. The damping of the CMB angular power spectrum tail at small scales (suggested by Crossing functions) is clearly visible in the inset plot. }
\end{figure*}

In Fig.~\ref{fig:clttplanck} we plot the Planck data in different frequencies and the best fit $\cl$'s obtained implementing different orders of Crossing function. As we could see in Fig.~\ref{fig:cf} the damping of the CMB tail at small scales is clearly visible here too. The theoretical models are plotted without the foreground nuisance parameters and the data points are 
plotted after the subtraction of foreground effects in each channel where we have calculated the foreground effects using the best fit nuisance parameters corresponding to the concordance 
model. Hence in practice the black line (corresponding to the best fit $\Lambda$CDM concordance model) should follow the data points completely since the subtracted foreground contributions correspond to this particular model. 

\begin{figure*}[!htb]
\begin{center} 
\resizebox{360pt}{240pt}{\includegraphics{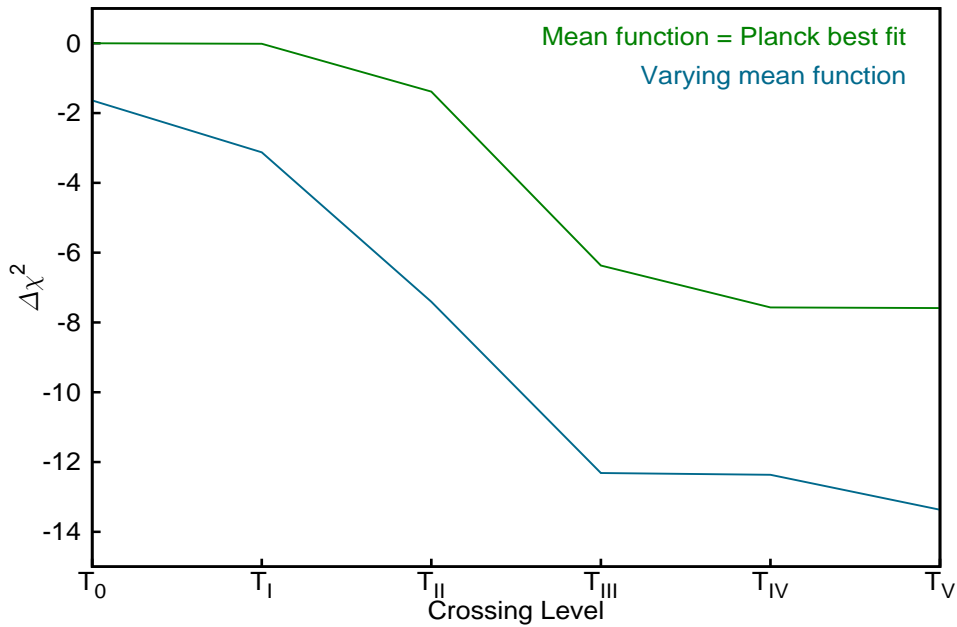}} 
\end{center}
\caption{\footnotesize\label{fig:chi2} $\Delta \chi^2$ with respect to the best
fit concordance model for different orders of Crossing function is plotted. Green line represent
the case that we use the best fit concordance model as a fixed mean function and blue line is for the case that 
we allow the parameters of the mean function to vary. Improvement in the likelihood for $T_{\rm III}$ case is particularly substantial and noticeable. 
}
\end{figure*}

In Fig.~\ref{fig:chi2} we plot the improvement in the likelihood to the Planck data with respect to the best fit concordance model for each level of Crossing function. Green line is for the case where the mean function is fixed to be the best fit concordance model and the blue line is for the case where we allow the parameters of the mean function (concordance model) to vary. 

Here we have to clarify few issues.  
The zero'th hyperparameter $C_0$, as the only hyperparameter of $T_0$ (in Eq.~\ref{eq:Crossing-function}) and also present in the higher orders of Crossing function, only shift the realizations of the mean function up and down. $C_0$ can basically be identified with the spectral amplitude of the primordial power spectrum (barring the lensing contribution) and hence no significant improvement is expected implementing $T_0$ Crossing function. We should note that there is a slight difference between effects of $C_0$ and the spectral amplitude on the overall shape of the angular power spectrum when we consider lensing. 
Spectral amplitude($A_{\rm s}$) changes the amplitude of the angular power spectrum and we include the lensing effects thereafter, while, $C_0$ shifts the whole angular power spectrum up and down after taking in to account the effect of lensing. 
Similarly the Crossing hyperparameter $C_1$, present in $T_{I}$ and higher orders of Crossing function, allows the whole spectrum to have a tilt and this can be thought to have a similar effect as of the $n_{\rm s}$ (within the cosmological parameters) on the form of angular power spectrum (barring the lensing contribution). Note that $n_{\rm s}$ is a tilt in wavenumber $k$ in the form of the primordial spectrum, which is not exactly the same as a tilt in $\ell$.   


Starting from the second order Crossing function we allow considerable deviation from the concordance model and we expect significant improvements in the likelihood if 
there are additional features in the data beyond the expectations of the concordance model. 
This is in fact what we see considering second and third order Crossing functions. In particular considering third order Crossing function that practically includes 
two additional degrees of freedom ($C_2$ and $C_3$ represent the new degrees of freedom, where $C_0$ and $C_1$ are similar to $A_{\rm s}$ and $n_{\rm s}$) we can get $12.5$ improvement in the $\chi^2$ that is 
substantial. This is the reason that the concordance model is pushed outside $2\sigma$ CL in the marginalized Crossing hyperparameter space. 
It is also interesting to see that considering the best fit concordance model as the single realization of the mean function, or varying the mean function parameters, can affect the improvements
in the likelihood substantially. This indicates that allowing the additional flexibilities from the Crossing functions, the best fit concordance model to the Planck data 
might not be the best mean function realization.  
 
To obtain the minimum $\chi^2$ values we use Powell's BOBYQA method of iterative minimization~\cite{powell} and detailed results are provided 
in Table~\ref{tab:chi2} and~\ref{tab:chi2-varybg}  The $\chi^2$ obtained from low-$\ell$ and high-$\ell$ are provided separately for comparison. Table \ref{tab:chi2} 
provides the results for the case of the fixed single mean function (best fit concordance model) and ~Table \ref{tab:chi2-varybg} is for the case that we allow parameters of the mean function 
to vary.

Considering the best fit concordance model as a fixed mean function, Crossing functions try to find optimal forms fitting the data better and we can 
see that it is not easy for these Crossing functions to improve the fit to the data at low and high multipoles simultaneously. However, allowing the mean
function also to vary (varying parameters of the mean function), there are combinations that can fit the data substantially better than the best fit concordance model at both low and high multipoles
(look at $T_3$ and $T_4$ in Table \ref{tab:chi2-varybg}).

Form of the Crossing functions (that their implementation can significantly improve the likelihood to the Planck data) suggest that there 
might be deviations from the concordance model at both low and high multipoles. In fact there are lack of power in the data at both 
low-$\ell$ and high-$\ell$ in comparison to the best fit concordance model as we can see in lower panels of Fig~\ref{fig:cf}. While these inconsistencies can
be due to statistical fluctuations in the data, to explain these deviations one may also consider some extensions to the concordance model. 
In fact there have been some attempts to explain the inconsistency in the derived cosmological parameters from Planck and other cosmological surveys by introducing new physics in the neutrino sector~\cite{Wyman:2013lza,Hamann:2013iba,Battye:2013xqa} or departure from cosmological constant as dark energy~\cite{Rest:2013bya}.
Some modifications in the foreground modeling can possibly resolve the issue too (the deviation at high-$\ell$) and this can be an attractive option if one does not want to introduce new physics to explain the data. As an always-present possibility, one can think of systematics in the data.


\begin{table*}[t]  
\begin{center}
  \hspace*{0.0cm}\begin{tabular}
  {|c|c|c|c|c|c|c|c|} \hline 
  Data & $\Lambda$CDM  & $T_{\rm II}$ &$T_{\rm III}$ & $T_{\rm IV}$ & $T_{\rm V}$ \\ \hline    

  {Planck low-$\ell$ ($\ell$=2-49)}	&  -6.3 &  -6.8 & -5.2 &-5.8 &-5.7 \\  \hline

  {Planck high-$\ell$ ($\ell$=50-2500)} & 7794.9 & 7794 & 7787.4&7786.8&7786.7\\  \hline

  {Total}		& 7788.6 & 7787.2& 7782.2&7781&7781\\  \hline 
  {$\chi^2_{\rm Model}$ - $\chi^2_{\rm \Lambda CDM}$} & - &-1.4 &-6.4&-7.6&-7.6 \\ \hline

  \end{tabular}
  \end{center}
  \caption{~\label{tab:chi2} The minimum $\chi^2$ values implementing different orders of Crossing function are tabulated. The $\chi^2$ values for high-$\ell$ and low-$\ell$ are provided separately. Here the best fit concordance model is used as a fixed mean function.
}
\end{table*}

\begin{table*}[t]  
\begin{center}
  \hspace*{0.0cm}\begin{tabular}
  {|c|c|c|c|c|c|c|c|} \hline 
  Data & $\Lambda$CDM  &$T_{\rm 0}$ &$T_{\rm I}$& $T_{\rm II}$ & $T_{\rm III}$ & $T_{\rm IV}$ & $T_{\rm V}$ \\ \hline    

  {Planck low-$\ell$ ($\ell$=2-49)}	  &  -6.3  & -7    &  -8.5   &  -8.6 & -9.8   &-9.7    &-9.7 \\  \hline

  {Planck high-$\ell$ ($\ell$=50-2500)}   & 7794.9 & 7793.8& 7793.8  & 7789.6& 7785.9 & 7785.7 &7784.7\\  \hline

  {Total}		                  & 7788.6 & 7786.8&  7785.3 & 7781  & 7776.1 & 7776&7775\\  \hline 
  {$\chi^2_{\rm Model}$ - $\chi^2_{\rm \Lambda CDM}$} &- &-1.8 & -3.3 &-7.6 &-12.5&-12.6&-13.6 \\ \hline

  \end{tabular}
  \end{center}
  \caption{~\label{tab:chi2-varybg} The minimum $\chi^2$ values implementing different orders of 
  Crossing functions are tabulated. The $\chi^2$ values for high-$\ell$ and low-$\ell$ are provided separately. Here we allowed the parameters of the mean function to vary.}
\end{table*}

\begin{figure*}[!htb]

\begin{center} 

\resizebox{140pt}{120pt}{\includegraphics{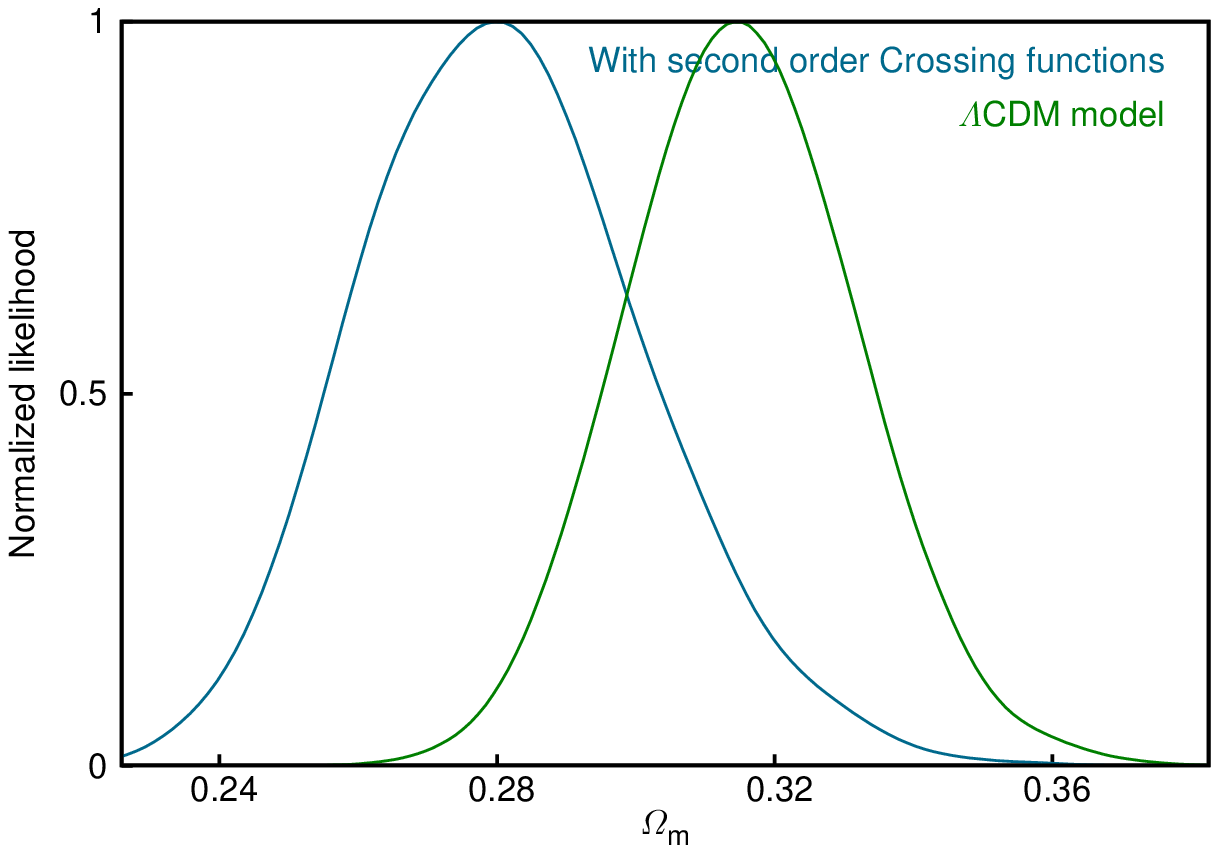}} 
\resizebox{140pt}{120pt}{\includegraphics{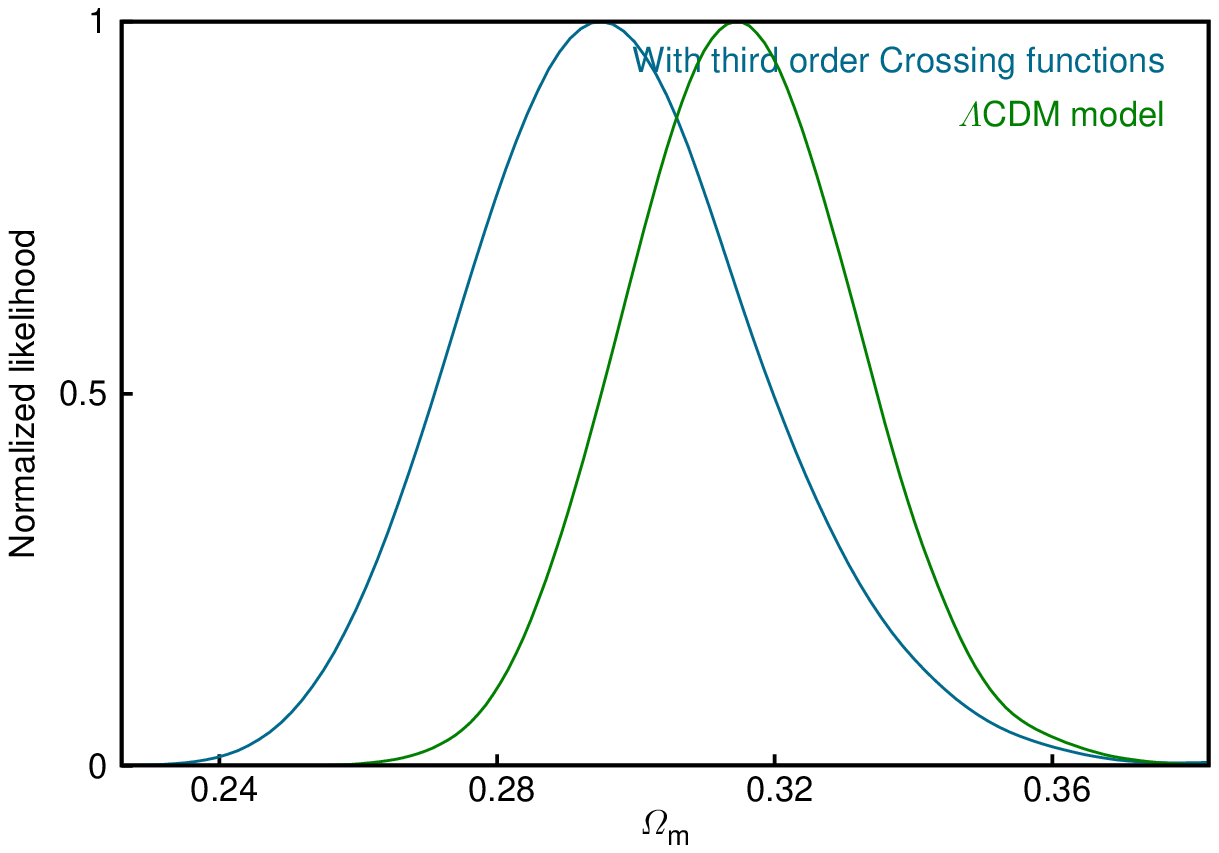}} 
\resizebox{140pt}{120pt}{\includegraphics{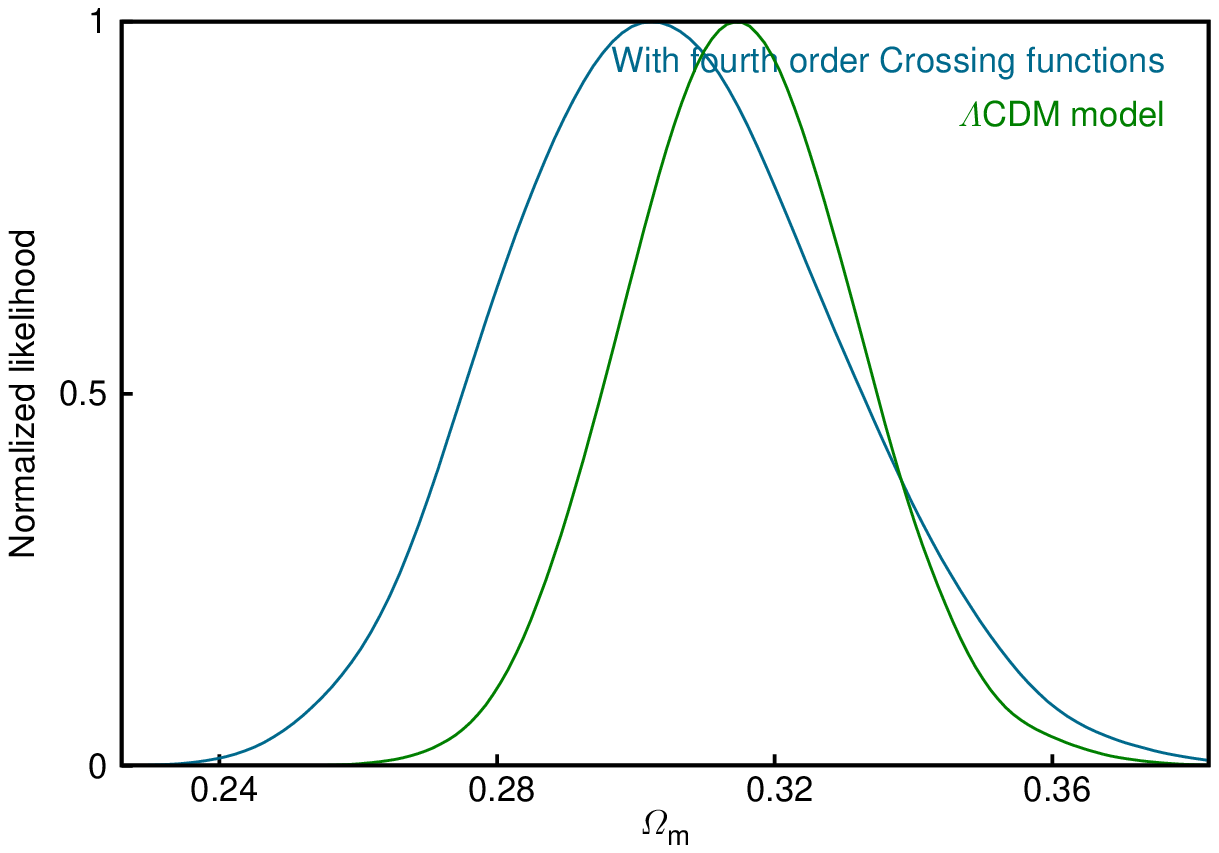}} 

\resizebox{140pt}{120pt}{\includegraphics{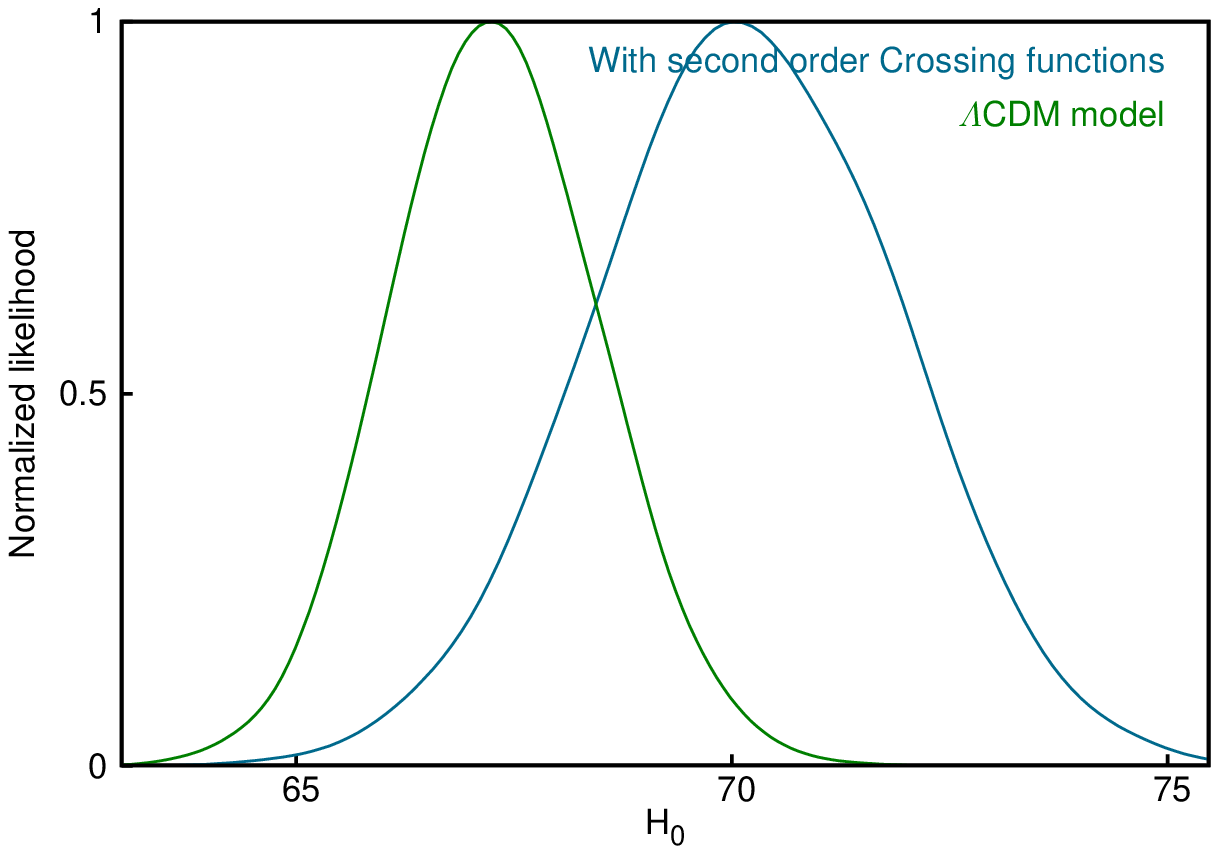}} 
\resizebox{140pt}{120pt}{\includegraphics{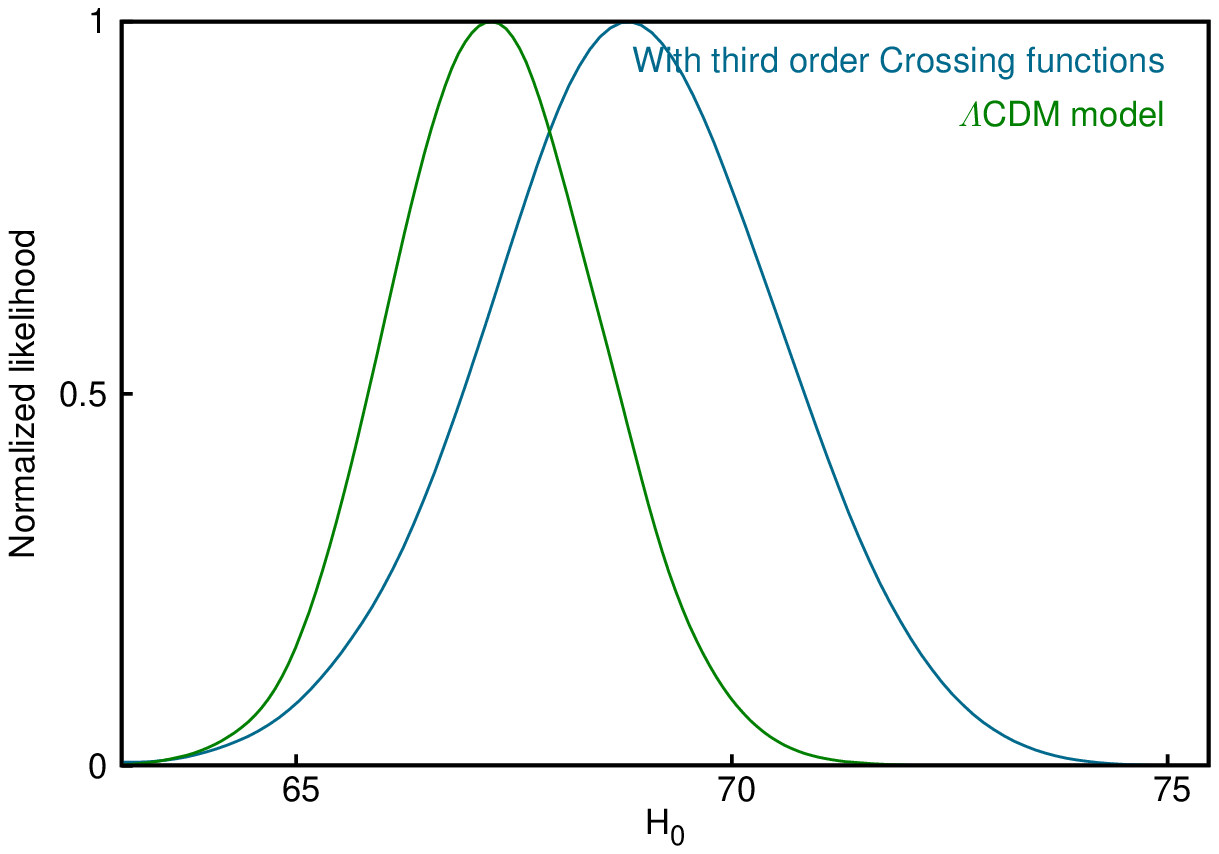}} 
\resizebox{140pt}{120pt}{\includegraphics{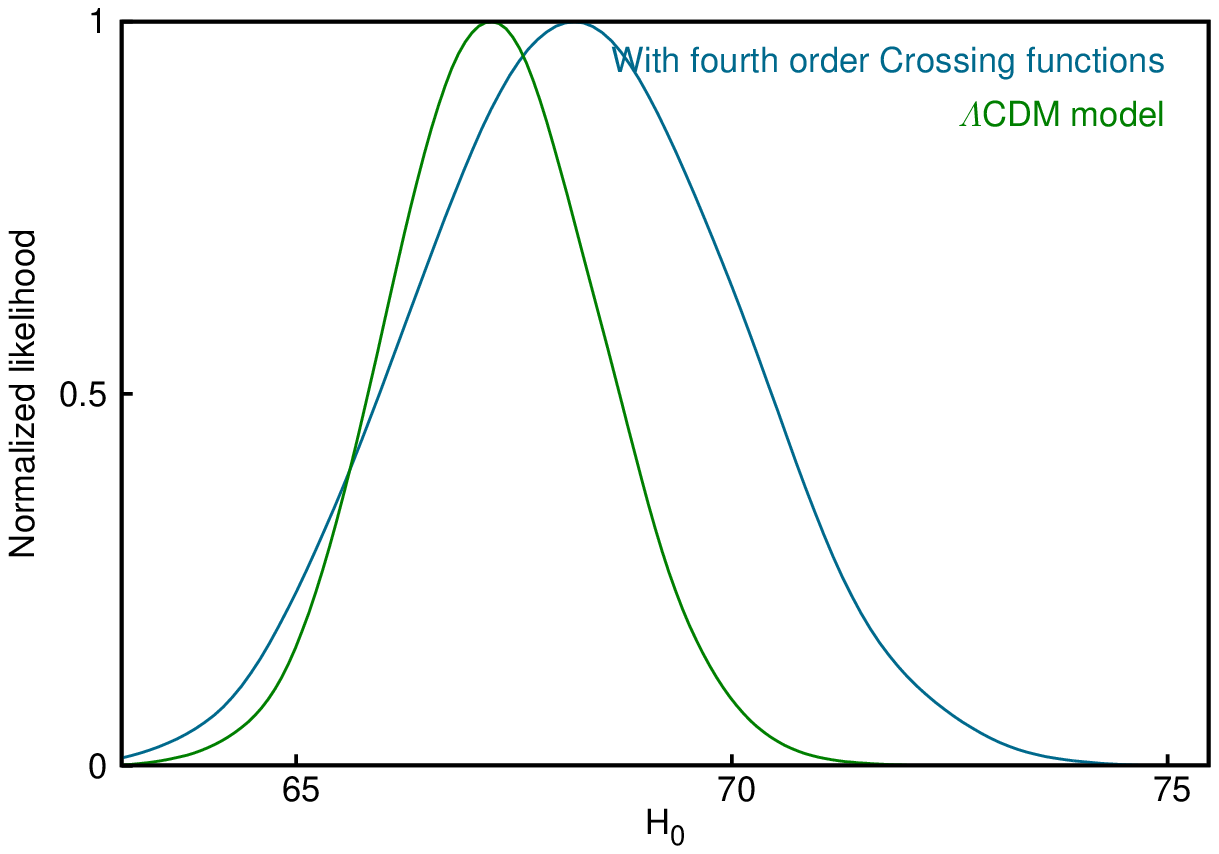}}

\resizebox{170pt}{140pt}{\includegraphics{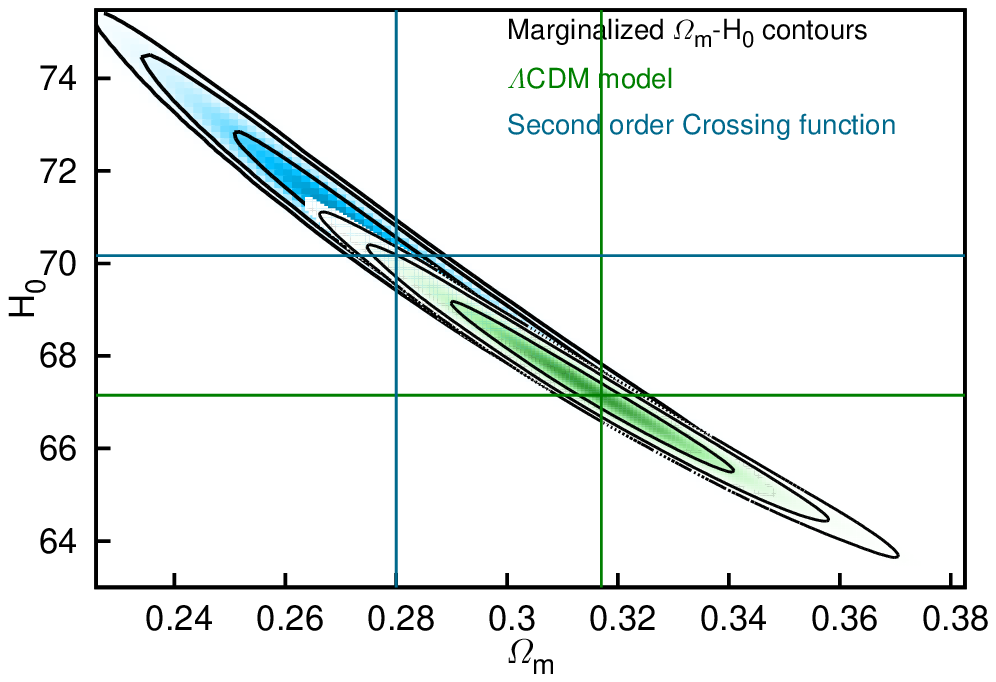}} 
\hskip -40 pt\resizebox{170pt}{140pt}{\includegraphics{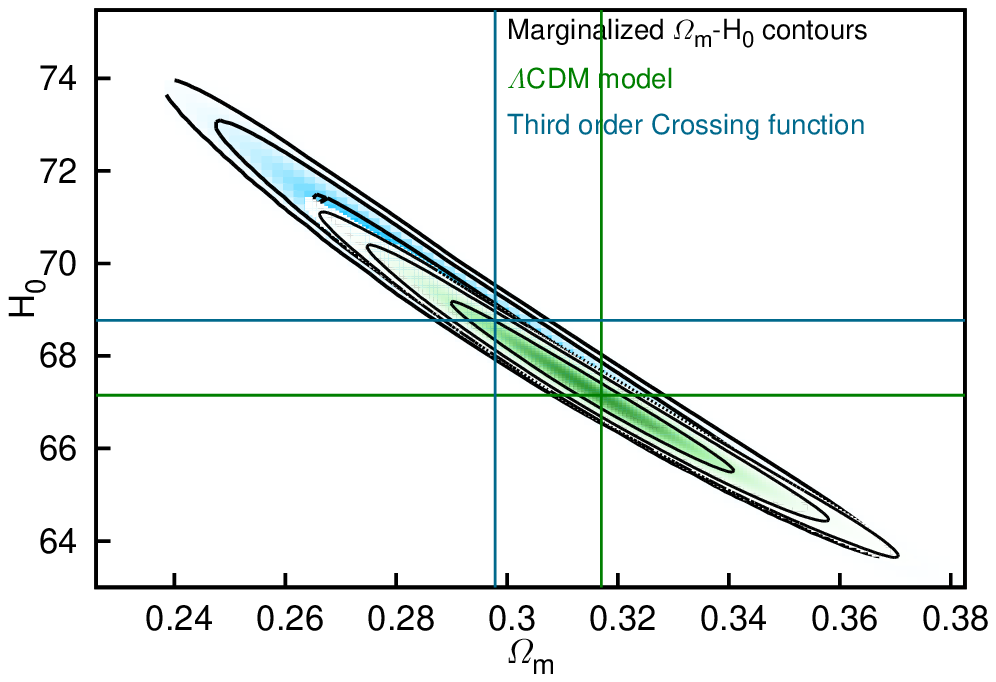}} 
\hskip -40 pt\resizebox{170pt}{140pt}{\includegraphics{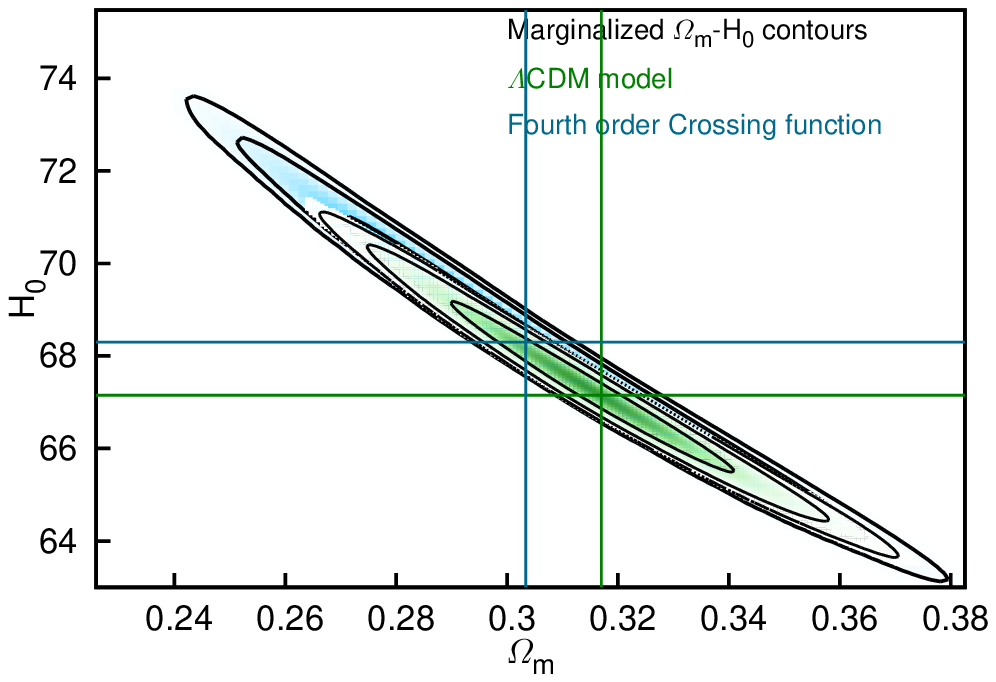}} 

\end{center}
\caption{~\label{fig:ommh0} 1 dimensional probability distribution functions and 2 dimensional 
contours of $\Omega_{\rm m}$ and $H_{0}$ (representing the parameters of the mean function) implementing Crossing 
functions of different order are plotted . Green color represent the case of the concordance model and the blue
color represent the results of our analysis using Crossing functions. The perpendicular lines show the best fit points 
for each contour plot. Implementing Crossing functions, data prefers lower matter density and higher Hubble parameter
for the concordance model.}
 \end{figure*}

Finally, at the end of this section we discuss briefly about the parameters of the mean function. In our analysis, 
allowing the parameters of the mean function to vary we get different forms of the mean function. Similarly, varying the hyperparameters of the Crossing function we get different forms of the Crossing function. Intermixing all these different realizations from the mean function and the Crossing function we fit the data and 
find the best combinations. While the form of the Crossing functions matter in order to understand what sort 
of deviation from the concordance model might improve the likelihood to the data, it is also important 
to see what forms of the mean function (which points in the parameter space of the mean function) are preferred by the data. 

In Fig.~\ref{fig:ommh0} we have plotted the marginalized probabilities of two  
cosmological parameters $\Omega_{\rm m}$ and $H_0$ considering Crossing functions of second,
third and fourth order. In the first and second row 1-D 
marginalized probability distribution of $\Omega_{\rm m}$  and $H_0$ are plotted. In the third row we plot 2-D 
marginalized contours of $\Omega_{\rm m}$  and $H_0$. The probability distributions and contours with blue color represent the 
results of our analysis considering Crossing functions and the green color represent the results for the case of the concordance model. 
It is interesting to see that considering Crossing functions the data prefers lower matter density and higher Hubble parameter for the 
concordance model. This may hint towards a fact that if we can provide with a modified form of foreground modeling that can explain the 
lack of power at high-$\ell$, the estimated values of the cosmological parameters from Planck can become more consistent with the values
reported by other surveys and analysis~\cite{Efstathiou:2013via}.

\section{Conclusions}\label{sec:conclusions}
In this paper we have implemented the Crossing statistic to test the consistency of the concordance model of cosmology with the Planck CMB data. We modeled the deviation from the concordance model using different orders of the Crossing function. Statistical significance of the deviation from the concordance model was estimated by deriving the confidence limits of the Crossing hyperparameters while we marginalized over parameters of the mean function. Crossing hyperparameters perform as discriminators between true and false assumed models (mean functions). If a Crossing function (that is a function of its hyperparameters) is derived to be nearly one fitting the data, it shows that the assumed mean function has a good consistency to the data and no modification is required to fit the data better. However, if Crossing functions are derived to be significantly different from one (constant at one), it reflects that the data suggests some features beyond flexibility of the assumed mean function. In our analysis, concordance model of cosmology provided the pool of mean function realizations (for different set of parameters) and we used Chebyshev polynomials up to fifth order as the Crossing functions. While using second order Crossing function shows proper consistency between the concordance model and Planck CMB data (Planck data was used through the likelihood code provided by the survey), we have noticed substantial improvement in the likelihood using third order Crossing function. Using Crossing functions of third and fourth order, the concordance model is pushed outside the $2\sigma$ confidence limits of the Crossing hyperparameters. This is a substantial deviation and needs proper investigation. Form of the best fit Crossing functions suggest that there is a lack of power in the data at both low-$\ell$ and high-$\ell$ in comparison to the concordance model. In fact smooth damping of the tail of the theoretical $\cl$ can improve the likelihood to the data significantly. 

There might be different explanations for this deviation. The $2\sim3\sigma$ inconsistency between the Planck data and concordance model might be due to statistical fluctuations. As another possibility, the form of the best fit Crossing functions suggests that a modified foreground modeling might be helpful to resolve this inconsistency at high-$\ell$. We have also noticed that considering Crossing functions, the data prefer lower matter density and higher Hubble parameter for the concordance model (that provides the pool of mean function realizations) with respect to the values derived by fitting the concordance model directly to the Planck data. This shows that resolving the issue at high-$\ell$ can also resolve the apparent inconsistency between the derived cosmological parameters from Planck and WMAP. Third possibility is that we are observing a real physical effect suggesting an extension to the concordance model. Last but not least possibility is that there might be some systematics in the data.

At the end we should note that we have only used publicly available Planck likelihood code in our analysis that includes all frequency channels. Performing the analysis on the individual frequencies can possibly help to understand which channel is more responsible for the required damping in the tail of the spectrum. This will be addressed in a separate analysis where we can test the consistency of the data from different channels including Planck polarization results and the revised likelihood code. In fact shortly before we finalize this work, there was a revised version of Planck XVI paper~\cite{Ade:2013zuv} where they have mentioned a small systematic in the $217 \times 217$ spectrum (apparently due to incomplete removal of $4$ K cooler lines). This systematic seems to be responsible for the feature at $\ell \sim 1800$ in the angular power spectrum. This hints that what we found as a significant damping (deficit of power in the data) at the high-$\ell$ might be due to the same systematic since our Crossing functions are not able to fit localized features. In other words, to fit the data better Crossing functions might have damped the tail of the spectrum to compensate the significant feature at $\ell \sim 1800$. While this looks like a possibility, more investigation is required and we have to wait for the new modified likelihood code and polarization data (where the effects of systematics are under control) to see if our results persist.




\section*{Acknowledgments}
We would like to thank Jan Hamann, Eiichiro Komatsu, Eric Linder and Tarun Souradeep for useful discussions. We wish also to acknowledge support from the Korea Ministry of 
Education, Science and Technology, Gyeongsangbuk-Do and Pohang City for Independent Junior Research Groups at 
the Asia Pacific Center for Theoretical Physics.  A.S. would
like to acknowledge the support of the National Research Foundation of Korea (NRF-2013R1A1A2013795).We also acknowledge the use of publicly available CAMB and CosmoMC in our analysis.

\end{document}